# Meta-Learned Per-Instance Algorithm Selection in Scholarly Recommender Systems

Andrew Collins
ADAPT Centre, School of
Computer Science and Statistics,
Trinity College Dublin, Ireland
ancollin@tcd.ie

Joeran Beel
ADAPT Centre, School of
Computer Science and Statistics,
Trinity College Dublin, Ireland
beelj@tcd.ie

## ABSTRACT

The effectiveness of recommender system algorithms varies in different real-world scenarios. It is difficult to choose a best algorithm for a scenario due to the quantity of algorithms available, and because of their varying performances. Furthermore, it is not possible to choose one single algorithm that will work optimally for all recommendation requests. We apply meta-learning to this problem of algorithm selection for scholarly article recommendation. We train a random forest, gradient boosting machine, and generalized linear model, to predict a best-algorithm from a pool of content similarity-based algorithms. We evaluate our approach on an offline dataset for scholarly article recommendation and attempt to predict the best algorithm per-instance. The best meta-learning model achieved an average increase in F1 of 88% when compared to the average F1 of all base-algorithms (F1; 0.0708 vs 0.0376) and was significantly able to correctly select each base-algorithm (Paired t-test; $p < 0.1$). The meta-learner had a 3% higher F1 when compared to the single-best base-algorithm (F1; 0.0739 vs 0.0717). We further perform an online evaluation of our approach, conducting an A/B test through our recommender-as-a-service platform Mr. DLib. We deliver 148K recommendations to users between January and March 2019. User engagement was significantly increased for recommendations generated using our meta-learning approach when compared to a random selection of algorithm (Click-through rate (CTR); 0.51% vs. 0.44%, Chi-Squared test; $p < 0.1$), however our approach did not produce a higher CTR than the best algorithm alone (CTR; MoreLikeThis (Title): 0.58%).

## 1 Introduction

Recommendation algorithm performance varies in different scenarios. A reliable intuition about what algorithms are best suited to a given scenario can be elusive even to recommender system experts [17], and it is generally accepted that manual experimentation is required [4]. Correctly choosing an optimal, single, algorithm however, will reduce the effectiveness of the system overall [2][10]. It is not possible to choose one algorithm that works optimally for all recommendation requests.

Within real-world recommendation scenarios, algorithm performance is unpredictable. For example, a near-to-online evaluation of recommendation algorithms across 6 online news websites showed that the performance of approaches was inconsistent [3]; a "most popular" algorithm performed best on one website (*cio.de*; precision: 0.56) and was the worst on another (*ksta.de*; precision: 0.01). In the domain of scholarly article recommendation, we performed an online evaluation of 33.5M recommendations delivered across multiple applications and found similarly inconsistent algorithm performances [8]; the best performing algorithm in one application (Document embeddings; Click-through rate (CTR): 0.21%) was the worst performing in another (CTR: 0.02%).

Algorithm performance is also unpredictable when considered at a per-instance level. In another previous evaluation we found that, for example, the overall-worst performing collaborative filtering algorithm for prediction in MovieLens datasets was frequently more accurate than all other algorithms for each user-item rating prediction [9]. If you can accurately predict when such an algorithm is optimal for each user-item prediction, significant gains in recommender system performance can be achieved (e.g., picking an optimal algorithm per-instance improves RMSE by 25.5% over the overall-best algorithm in this case [9]).

We face these challenges of variable and inconsistent algorithm performance as operators of the scholarly recommender-as-a-service Mr. DLib. We work with diverse partners who each have different users, different websites or applications, different corpora, etc. It is unknown how algorithms will perform for new partners, or for each recommendation request that is made, and it is not sufficient to choose algorithms that have performed well for previous partners [8].

Meta-learning for algorithm selection aims to predict the best algorithm to use in a given situation. It does this by learning the relationship between characteristics of data, and the performance of algorithms for that data [19][25]. For example, 'sparsity' is a characteristic of ratings data (a *meta-feature*), and it might be learned that collaborative filtering algorithms are more effective on non-sparse datasets. Meta-learning is useful when distinct algorithms perform differently in various situations, when those situations that can be characterized numerically, and when this performance variation can be learned.

In this paper we apply meta-learning for algorithm selection to the task of scholarly article recommendation. We aim to select the best algorithm for each data instance, and for each recommendation request received. We evaluate several approaches using a gold-standard offline dataset and deploy the most promising approach to a live recommender system for online evaluation.

## 2 Related Work

Meta-learning for algorithm selection in recommender systems is typically used to predict a single best-algorithm for entire datasets [10]. Several offline evaluations of algorithm selection using meta-learning for recommender systems exist in which best-algorithms are predicted per-dataset, for example, Cunha et al. [11, 13], Romero et al. [22], Matuszyk and Spiliopoulou [21]. Furthermore, tools exist to automatically evaluate algorithm pools for recommender systems at a per-dataset level, such as librec-auto [20]. See Cunha et al. [12] for a survey of recommender system-related algorithm selection literature and Smith-Miles [23] for a wider survey of meta-learning for algorithm selection.

Per-instance meta-learning is well explored in other fields; see Kotthoff [18] for a comprehensive overview of per-instance meta-learning and its successful application to combinatorial search problems for example.

There are not many examples of meta-learning being used for recommender systems at lower levels of granularity than per-dataset, that is, predicting the best algorithm for subsets of data, or per-instance.

Ekstrand and Riedl use a meta-learner to choose the best algorithm per-user, from a small pool of algorithms, using a MovieLens dataset [15]. They predict an algorithm using two meta-features: the number of ratings by a user, and the variance of a user's ratings. Their meta-learner performs slightly worse than the overall-best single algorithm (RMSE; 0.78 vs 0.77).

**Table 1. An illustration of the dataset used to train our meta-learners. The target is a categorical variable that indicates the best algorithm and search fields for this instance according to an F1-score**

| Researcher ID | Doc. ID | Collection ID | Title length (chars) | Title length (words) | Best Algorithm (F1) |
|---|---|---|---|---|---|
| y1 | 1062 | P00 | 104 | 14 | MLT:title,abstract |
| y1 | 1064 | P00 | 45 | 5 | MLT:title,abstract |
| y1 | 1029 | P03 | 88 | 11 | StandardQuery:title,abstract |
| … | … | … | … | … | … |
| y10 | 1053 | P04 | 61 | 8 | MLT:title,abstract |

Collins et al. [10] predict the best algorithm per-instance from a pool of collaborative filtering algorithms using two MovieLens datasets (100K and 1M). Their meta-learners were unable to discriminate algorithm performance adequately (RMSE, 100K: 0.973; 1M: 0.908) and performed 2-3% worse than overall-best single algorithm SVD++ (RMSE, 100K: 0.942; 1M: 0.887).

Edenhofer et al. [14] use per-instance meta-learning on an offline dataset and find that a gradient boosting model improves recommender system effectiveness slightly when compared to the overall-best algorithm.

There are no *online* evaluations of instance-based meta-learners for recommender systems to the best of our knowledge[1]. The effectiveness of meta-learning for algorithm selection in live recommender systems is not known, as far as we are aware.

## 3 Methodology

We hypothesize that there might be a relationship between text-based attributes of recommendation requests to a scholarly article recommender system, and the performances of content-similarity algorithms for those requests. For example, the length of a text query may alter a TF-IDF-based algorithm's effectiveness in a recommendation scenario, as shown in other domains [7]. If such a relationship can be learned by a meta-learner, then we expect that algorithm selection can improve engagement with scholarly article recommendations through the use of a more appropriate algorithm for a given request.

To evaluate meta-learning models' abilities to learn such a relationship and increase user engagement we perform a two-stage evaluation. We first identify candidate meta-learning models via an offline evaluation. We then deploy any promising model to a live recommender system for an online evaluation.

### 3.1. Offline Evaluation

We approach our offline evaluation as a classification task. We use the Scholarly Paper Recommendation Dataset [24].

---

[1] A comprehensive summary of algorithm selection literature is available here: https://larskotthoff.github.io/assurvey/

This dataset contains 597 papers from a corpus of scholarly publications about Computational Linguistics (ACL Anthology Reference Corpus [2]). The interests of 28 researchers are described in the dataset, and these researchers have manually indicated what papers within the corpus are relevant for them. A median of 30 relevant papers per-researcher are indicated. We import the titles and abstracts for all 597 documents into Solr[3].

We use two content similarity-based search algorithms built into Solr for our evaluation. For each document, in each researcher's repository, we perform four queries as follows:

1) Two searches using Solr's *standard query parser*. We use the title from each row of the dataset and search all documents in the corpus on either their title fields, or, search all documents using both their title and abstract fields.

2) Two searches using Solr's MoreLikeThis (MLT) class. MoreLikeThis constructs a term vector from either the title field, or both the title and abstract fields, and returns similar documents.

Both approaches use TF-IDF and a scoring formula similar to cosine similarity[4] to rank results. The substantive difference between these two approaches is that the standard query parser will only use the title from each instance to find results, and MLT may use the abstract also.

We compare the documents returned from each of the four queries to the remaining documents marked as relevant by that researcher. We rank the algorithms according to the F1 score of the retrieved results and note the best performing algorithm. We also derive the length of the querying document's title in characters and words, and note the collection that the querying document is in. If all algorithms return zero results for a row, the row is removed. This results in a dataset with 750 instances (**Table 1**).

We aim to learn a relationship between the performance of the base-algorithms and characteristics of the instance data. To do this, we train and evaluate three models: a random forest, generalized linear model, and gradient boosting machine[5]. As features we use the Collection ID[6], Title Length (characters), and Title Length (words). The training target (meta-*target*) for each model is the label of the actual best algorithm per-instance according to F1 (i.e., the Best Algorithm column in **Table 1**). The simple features that we use for this offline evaluation are representative of the limited features that we are able to use with some partners in our live recommender system.

---

[2] http://acl-arc.comp.nus.edu.sg/
[3] We use the Cleaned ACL ARC dataset for titles and abstracts, available here: https://web.eecs.umich.edu/~lahiri/acl_arc.html [5]
[4] https://lucene.apache.org/core/7_7_0/core/org/apache/lucene/search/similarities/TFIDFSimilarity.html
[5] We use H2O's implementations of these models

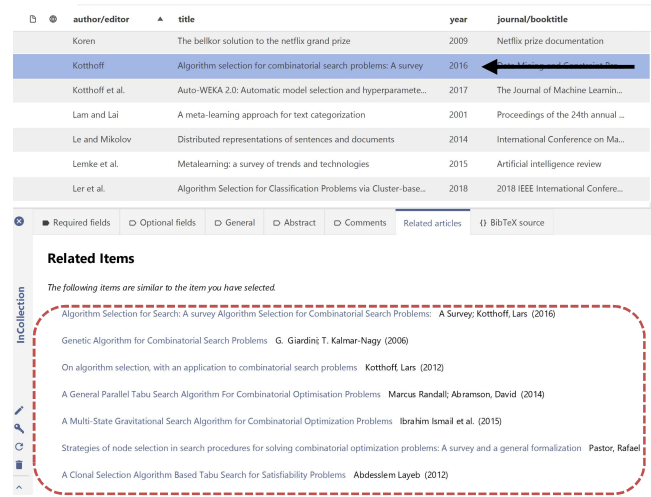

**Fig. 1. Recommendations generated by Mr. DLib and displayed in Jabref. Recommendations are generated for the currently selected repository item (indicated by a black arrow) and are displayed in a vertical list of up to 7 items (indicated with a dashed red box).**

We perform a 4-fold cross validation for each of these models. For each instance in the validation set from each fold, we predict the best algorithm and perform a Solr search using this predicted-best algorithm. We evaluate their results according to precision, recall, and F1.

We further evaluate each base-algorithm across each validation set. As a simple baseline we also use a randomly selected algorithm for each row. To illustrate the upper-bound of effectiveness that might be expected, we also use a "best-algorithm oracle" that uses the actual-best algorithm per-instance.

Our offline evaluation mirrors our online system, but with gold-standard data. 580 of the 597 documents in the small Scholarly Paper Recommendation Dataset are contained in the corpus used in our online evaluation.

### 3.2. Online Evaluation

Mr. DLib is a recommendation-as-a-service provider that delivers scholarly related-article recommendations for its partners. It is a white label of Darwin & Goliath [5]. Mr. DLib partners with Jabref, an open-source reference management software, and delivers recommendations to their users [16].

Mr. DLib in-part uses the same content-similarity based algorithms that we use in our offline evaluation to generate recommendations, specifically TF-IDF similarity using: Solr's standard query parser and a requesting document's title, and using Solr's MoreLikeThis on a requesting document that is known and in Mr. DLib's corpus.

---

[6] The collection ID contains a letter indicating the venue the item was published in along with the type of publication (proceedings/journal/workshop), and 2 numbers indicating year of publication. For a description of ACL's naming convention see: https://aclweb.org/anthology/info/contrib/

Table 2. Precision, recall, and F1 for each base-algorithm when selected arbitrarily, and for each base-algorithm when chosen by a meta-learner model (Random Forest, Generalized Linear Model, Gradient Boosting Machine). The highest precision, recall, and F1 for each base-algorithm is bolded. The overall per-instance meta-learner performance for each model is listed, with the best precision, recall and F1 also highlighted

| Algorithm | Precision | Recall | F1 |
|---|---|---|---|
| **MoreLikeThis (Title)** | | | |
| Arbitrarily Selected | 0.012 | 0.001 | 0.002 |
| Random Forest Meta-learner | **0.383** | **0.039** | **0.070** |
| Generalized Linear Model Meta-learner | 0.000 | 0.000 | 0.000 |
| Gradient Boosting Machine Meta-learner | 0.300 | 0.028 | 0.050 |
| **MoreLikeThis (Title, Abstract)** | | | |
| Arbitrarily Selected | 0.313 | 0.042 | 0.072 |
| Random Forest Meta-learner | **0.327** | 0.044 | **0.075** |
| Generalized Linear Model Meta-learner | 0.080 | 0.011 | 0.018 |
| Gradient Boosting Machine Meta-learner | 0.332 | **0.046** | 0.077 |
| **Standard Query Parser (Title)** | | | |
| Arbitrarily Selected | 0.117 | 0.022 | 0.035 |
| Random Forest Meta-learner | **0.191** | **0.047** | **0.072** |
| Generalized Linear Model Meta-learner | 0.032 | 0.005 | 0.008 |
| Gradient Boosting Machine Meta-learner | 0.142 | 0.034 | 0.052 |
| **Standard Query Parser (Title, Abstract)** | | | |
| Arbitrarily Selected | 0.123 | 0.027 | 0.041 |
| Random Forest Meta-learner | 0.215 | 0.040 | 0.066 |
| Generalized Linear Model Meta-learner | 0.056 | 0.012 | 0.019 |
| Gradient Boosting Machine Meta-learner | **0.218** | **0.051** | **0.077** |
| **Best Algorithm Oracle** | 0.370 | 0.060 | 0.098 |
| **Random Algorithm** | 0.137 | 0.024 | 0.038 |
| **Per-Instance Meta-learners Overall** | | | |
| Random Forest Meta-learner | **0.303** | **0.044** | **0.074** |
| Generalized Linear Model Meta-learner | 0.168 | 0.028 | 0.046 |
| Gradient Boosting Machine Meta-learner | 0.286 | 0.044 | 0.073 |

Mr. DLib makes recommendations to users of Jabref from 120M documents in the CORE [7] collection of open access research papers. Jabref makes a request for recommendations using the title of the currently selected document in the user's repository (**Fig. 1**). Related-articles are then recommended based on text and metadata from documents in the corpus, including the title and abstract. Recommendations are displayed to users in a vertical list of up to 7 items (**Fig. 1**).

We measure the effectiveness of each base-algorithm using click-through rate (**CTR**). This is the ratio of clicked recommendations to delivered recommendations. For example, if 1000 recommendations are delivered and 9 are clicked, this gives a CTR of $\frac{9}{1000} = 0.9\%$. We assume that if an algorithm is effective then, on average, users will interact with recommendations from this algorithm more frequently than recommendations from a less effective one. This will manifest as a higher CTR for the more effective algorithm.

We train and deploy the most effective meta-learning model from our offline evaluation. We use 2 months of user-interaction data to train this model, logged from November 2018 to January 2019. Our training set comprises recommendations that were clicked by users and describe each querying document's title length (words), title length (characters), and the hour of the day that a request was received. Recommendations are infrequently clicked and so our training set only includes ~1% of total recommendations delivered in this period. The target of our model is the label of the algorithm used to generate these previously clicked requests. Algorithms were selected with equal probability during the period in which training click-data was collected.

During the evaluation period, half of recommendation requests received by Mr. DLib were fulfilled using the predicted-best algorithm from our meta-learner. Unlike the offline evaluation, only one algorithm can be used for any specific recommendation request/instance in this live recommender system, therefore the remaining half of recommendation requests are fulfilled by a random selection of algorithm (MoreLikeThis search, standard query search) and search field (title, title and abstract). Mr. DLib only uses MoreLikeThis if a querying document is indexed in Solr [8]. In the case that a querying document is not in Mr. DLib's corpus, a fallback algorithm is used.

Our evaluation is based on recommendations delivered to users of Jabref between January 2019 and March 2019.

## 4 Results

### 4.1. Offline Evaluation Results

Results from our offline evaluation are shown in Table 2. For each base-algorithm, precision, recall and F1 are always higher for instances where the algorithm is selected by a meta-learner, versus when the base-algorithm is used alone (i.e., arbitrarily selected). The random forest achieves a significantly higher F1 (Paired t-test; p < 0.1) on a per-algorithm basis over arbitrary use of the base-algorithms (88% increase, average F1; 0.0708 vs 0.0376), followed by gradient boosting (71% increase, average F1; 0.0644 vs 0.0376). A generalized linear model produces a lower average F1 than the base-algorithms (70% decrease, average F1; 0.0114 vs 0.0376). The performance for each meta-learner overall, i.e., when the meta-learner selects what algorithm to use per-instance, is also listed in Table 2. Of the three meta-learning models, the random forest achieves the

---
[7] https://core.ac.uk/

[8] MoreLikeThis can also use an external resource to conduct a search. To minimize recommendation response time Mr. DLib does not use this mechanism

Algorithm Selection in Scholarly Recommender Systems

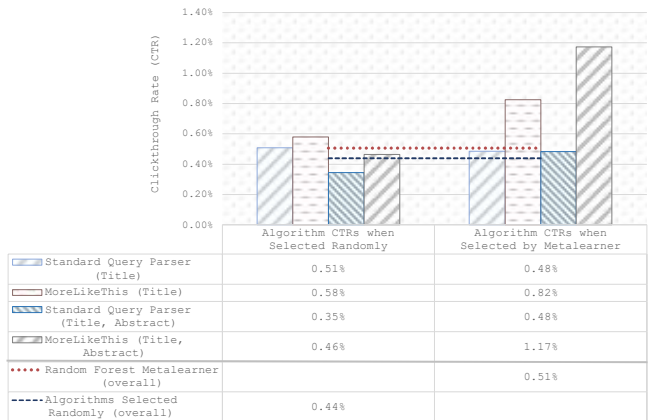

Fig. 2. Clickthrough rates for each base-algorithm, when used randomly, and when selected by the random forest meta-learner in Mr. DLib. A random forest meta-learner produces an overall 15.04% increase in CTR over a random selection of algorithm but is less effective than the best individual algorithm, evaluated on 148,088 recommendations.

highest precision, recall, and F1. The overall-F1 of the random forest is 3% higher than the overall-best base-algorithm (F1; 0.0739 vs 0.0717 for MoreLikeThis (Title, Abstract)) indicating the meta-learner is effective in selecting algorithms appropriately.

The upper bound on performance for these base-algorithms is indicated by the 'best algorithm oracle', which achieves a 32% higher F1 than the best meta-learner

### 4.2. Online Evaluation Results

Mr. DLib delivered 148,088 recommendations to users during the evaluation period, using the base-algorithms selected randomly, and using base-algorithms as selected by our random forest meta-learner. Overall there were 719 clicks upon recommendations, giving a total average click-through rate of 0.49%. This average click-through rate seems low but is consistent with previous large-scale evaluations that we have performed comprising 100M recommendations [6].

Results from our online evaluation are shown in Fig. 2. Recommendations from three of the four base-algorithms achieve a higher CTR when this base-algorithm is selected by the random forest meta-learner. MoreLikeThis (Title and Abstract) sees a ~150% increase in CTR when chosen by the meta-learner. Overall, the random forest meta-learner results in a 15.04% increase in CTR over a random selection of algorithms (CTR; 0.51% vs. 0.44%, Chi-Squared test; $p < 0.1$). However, the meta-learner's overall CTR is not higher than the best two algorithms individually (CTR; Standard Query Parser (Title): 0.51%, MoreLikeThis (Title): 0.58%).

## 5 Discussion and Conclusion

Our online evaluation shows that a random forest meta-learner, using a requesting document's title length in words and characters and the hour of the day that a request was received as meta-features, is less effective than MoreLikeThis (Title) alone (CTR; 0.58% vs 0.51%) and is only equally effective as the Standard Query Parser using just the Title (**Fig. 2**). Based on this evaluation, MoreLikeThis (Title) should be used instead of this meta-learning approach, or any of the other algorithms examined. However, although the meta-learner's effectiveness is not better than the overall best algorithm, the results are encouraging in that the meta-learner is, to some extent, capable of learning when an algorithm performs best. This is indicated by the meta-learner's higher effectiveness when compared to a random selection of algorithm. It is furthermore encouraging that these results are based on a small number of simple meta-features, and that the training set considers only one simple form of implicit feedback, i.e., clicks, in order to predict a best algorithm.

In contrast to this online evaluation, our offline evaluation showed that algorithm selection *was* effective with the dataset we used, that is, with a subset of Mr. DLib's main corpus along with gold standard indications of recommendation relevance. Each individual algorithm was always more effective when chosen by a meta-learner than when used arbitrarily. Furthermore, the overall F1 for the random forest and gradient boosting machine meta-learners was higher than any individual algorithm. The discrepancy between our offline and online results highlights the need to examine approaches in a live setting.

We have evaluated these algorithms previously [8] and found a significant variation in effectiveness across recommendation scenarios, that is, in different applications, for different users. The rank of algorithms according to their effectiveness also differed, with the best algorithm in one scenario being the worst in another, and vice versa. These aberrations occur even with common corpora. The single-best algorithm could therefore not be assumed without such an online evaluation. We feel that the meta-learning approach outlined is a simple alternative to a lengthy evaluation, or an arbitrary/random choice of algorithm.

To the best of our knowledge, this is the first online evaluation of meta-learning for algorithm selection in a scholarly recommender system. We hope that these results can be improved upon. Further work includes the use of a more substantial dataset for offline evaluation[9]. More discriminative meta-features should be evaluated, such as text features that correspond to a predictable performance for retrieval methods, e.g., average query IDF [1]. Furthermore, it will be necessary to examine the effectiveness of algorithm selection not only per-instance, but also across multiple scenarios and partners.

---

[9] E.g., the large Scholarly Paper Recommendation Dataset. It contains 100,531 papers and lists the interests of 50 researchers [22]. Currently this dataset is not complete and does not include reference information for papers.

**ACKNOWLEDGMENTS**

This publication has emanated from research conducted with the financial support of Science Foundation Ireland (SFI) under Grant Number 13/RC/2106.